\documentclass[11pt]{article}
\usepackage{mathrsfs,amsmath,amsbsy,amssymb,amsthm}

\def\be{\begin{equation}}
\def\ee{\end{equation}}
\def\bea{\begin{eqnarray}}
\def\eea{\end{eqnarray}}
\def\O[#1]{{\cal O}(\Omega^{#1})}

\def\nn{\nonumber \\}

\def\mc{\mathcal}

\DeclareTextFontCommand{\textwasy}{\wasyfamily}
\def \wasyfamily{\fontencoding{U}\fontfamily{wasy}\selectfont}
\def \thorn{{\wasyfamily\char105}}
\DeclareTextCommand{\dh}{OT1}{{\wasyfamily\char107}}
\newcommand{\tho}{{\textrm\thorn}}
\renewcommand{\eth}{{\textrm{\dh}}}

\newcommand{\half}{{\textstyle{\frac{1}{2}}}}
\renewcommand{\nn}{{\nonumber}}

\newcommand{\kap}{{\kappa}}

\newcommand{\Om}{\Omega} 
\newcommand{\Ps}{\Psi}   
\newcommand{\Phis}{\Phi^\mathrm{S}} 
\newcommand{\Phia}{\Phi^\mathrm{A}} 

\newcommand{\lb}{{\ell}}
\newcommand{\nb}{{n}}
\newcommand{\mb}[1]{{m_{(#1)}}}

\newcommand{\M}[1]{{\stackrel{#1}{M}}}  

\title{Peeling of the Weyl tensor and gravitational radiation in higher dimensions}

\author{Mahdi Godazgar and Harvey S. Reall\\ {\footnotesize Department of Applied Mathematics and Theoretical Physics,  Centre for Mathematical Sciences, }\\ {\footnotesize Wilberforce Road, Cambridge CB3 0WA, UK}\\ {\footnotesize mmg31@cam.ac.uk, hsr1000@cam.ac.uk}}

\begin{document}

\maketitle

\begin{abstract}
The peeling behaviour of the Weyl tensor near null infinity is determined for an asymptotically flat higher dimensional spacetime. The result is qualitatively different from the peeling property in 4d. To leading order, the Weyl tensor is type N. The first subleading term is type II. The next term is algebraically general in 6 or more dimensions but in 5 dimensions another type N term appears before the algebraically general term. The Bondi energy flux is written in terms of ``Newman-Penrose" Weyl components.
\end{abstract}

\section{Introduction}

In 4 spacetime dimensions, the Weyl tensor of an asymptotically flat spacetime exhibits the ``peeling" property:
\be
 C_{\mu\nu\rho\sigma} = \lambda^{-1} C^{(N)}_{\mu\nu\rho\sigma} + 
\lambda^{-2} C^{(III)}_{\mu\nu\rho\sigma} + \lambda^{-3} C^{(II)}_{\mu\nu\rho\sigma} + \lambda^{-4} C^{(I)}_{\mu\nu\rho\sigma} +{\cal O}(\lambda^{-5})
\ee 
where indices $\mu,\nu,\ldots$ refer to a basis parallely transported along an outgoing null geodesic with affine parameter $\lambda$. In the first term, $C^{(N)}_{\mu\nu\rho\sigma}$ is a Weyl tensor of algebraic type N and the subsequent terms involve Weyl tensors of algebraic types III, II and I. The tangent to the geodesics is the repeated principal null direction for the type N, III and II terms, and a principal null direction for the type I term.

This result was originally derived using Bondi coordinates \cite{bondi, Sachs:1962wk}. In this approach, one assumes that the metric components can be expanded in inverse powers of a coordinate $r$. A more geometrical proof can be given using the definition of asymptotic flatness in terms of a conformal compactification \cite{pencom, geroch}. In this case, the result follows from the assumed smoothness of the unphysical spacetime. This smoothness assumption (or the assumption of an expansion in inverse powers of $r$) excludes some spacetimes in which radiation is present near spatial infinity. In this case, the peeling property is modified by an ${\cal O}(\lambda^{-4} \log \lambda)$ term \cite{damour,christodoulou} (see also \cite{chrusciel}).\footnote{We thank M. Dafermos for pointing out these references.}

In $d>4$ dimensions, a definition of asymptotic flatness at null infinity using conformal compactification is possible only for even $d$ \cite{Hollands:2003ie,Ishibashi:2007kb}. It has been shown that this definition is preserved by linearized metric perturbations arising from compactly supported initial data \cite{Hollands:2003ie}. It has also been argued that a vacuum spacetime satisfying this definition arises from initial data describing a small (but finite) perturbation of Minkowski spacetime that coincides with Schwarzschild initial data outside some compact set \cite{ChoquetBruhat:2006jc}. Just as in 4d, there are more general initial data that do not give a smooth null infinity \cite{Chrusciel:2010eb}.

For odd $d$, conformal compactification is unsatisfactory because the unphysical spacetime cannot be smooth in any radiating spacetime \cite{Hollands:2004ac}. Instead, one can follow the older approach of defining asymptotic flatness at null infinity using Bondi coordinates \cite{Tanabe:2009va,Tanabe:2011es}. In section \ref{sec:defs} we will weaken this definition slightly and demonstrate equivalence of the conformal and Bondi definitions for even $d$. For odd $d$, it remains to be shown that there exists an interesting class of initial data that gives rise to a spacetime which satisfies this definition. 

The goal of this paper is to determine how the Weyl tensor peels near null infinity in a spacetime satisfying one of the above definitions of asymptotic flatness. As just mentioned, at least for even $d$, this includes a large class of physically interesting spacetimes, but probably also excludes some physically interesting spacetimes. However, we can hope that in the latter case, just as in 4d, the peeling behaviour is modified only at a sufficiently high order that our result is still useful. 

Two previous papers have investigated peeling using the conformal approach to asymptotic flatness \cite{Pravdova:2005ey,Krtous:2006jh}. Both papers concluded that peeling is similar to the $d=4$ case. They started from the assumption that all components of the unphysical Weyl tensor decay at the same (unspecified) rate near null infinity. However, Ref. \cite{Hollands:2004ac} showed that this assumption is not true even for linearized perturbations of Minkowski spacetime, and argued that peeling should be qualitatively different for $d>4$. This is what we find.

In section \ref{sec:Weyl}, we determine the behaviour of the Weyl tensor near null infinity in a spacetime satisfying the ``Bondi definition" of asymptotic flatness (since this is valid for odd or even $d$ and equivalent to the conformal definition for even $d$). For $d \ge 6$, we find the following result:
\be
\label{peeling}
 C_{\mu\nu\rho\sigma} = \lambda^{-(d/2-1)} C^{(N)}_{\mu\nu\rho\sigma} + \lambda^{-d/2} C^{(II)}_{\mu\nu\rho\sigma} +\lambda^{-(d/2+1)} C^{(G)}_{\mu\nu\rho\sigma} + \ldots.
\ee 
Again $\lambda$ is an affine parameter along a null geodesic and $\mu,\nu,\ldots$ refer to a parallelly transported basis. The superscripts N,II,G refer to the higher-dimensional classification of the Weyl tensor developed in Ref. \cite{Coley:2004jv}, based on the concept of Weyl Aligned Null Directions (WANDs). A type N or type II Weyl tensor admits a ``multiple WAND", in this case it is the tangent to the geodesic. The type II term in (\ref{peeling}) is not the most general type II Weyl tensor: it obeys additional restrictions explained below. Type G denotes an algebraically general Weyl tensor. The ellipsis in (\ref{peeling}) denotes terms of order $\lambda^{-(d/2+2)}$ (even $d$) or $\lambda^{-(d/2+3/2)}$ (odd $d$).

For even $d$, the derivation of this result requires no more than the definition of asymptotic flatness. We do not use the Einstein equation, so this result is valid for any energy-momentum tensor consistent with asymptotic flatness.
For odd $d$, we need to use some additional information from the Einstein equation: a mild condition on the decay of the Ricci tensor near null infinity is required to eliminate a term of order $\lambda^{-(d/2+1/2)}$ from (\ref{peeling}). 

The case $d=5$ is exceptional. In this case, the Einstein equation no longer eliminates the term of order $\lambda^{-(d/2+1/2)}=\lambda^{-3}$. Instead, it fixes this term to be quadratic in the leading order metric perturbation and hence non-zero in any radiating spacetime. The result is that an additional type N term appears between the type II and type G terms:
\be
\label{peeling5d}
 C_{\mu\nu\rho\sigma} = \lambda^{-3/2} C^{(N)}_{\mu\nu\rho\sigma} + \lambda^{-5/2} C^{(II)}_{\mu\nu\rho\sigma} +  \lambda^{-3}   C^{(N)'}_{\mu\nu\rho\sigma} +  \lambda^{-7/2} C^{(G)}_{\mu\nu\rho\sigma} + {\cal O}(\lambda^{-4}). 
\ee 
The subleading type N term is distinct from the leading order type N term. The presence of this term can be attributed to the nonlinearity of the Einstein equation. For $d>5$, nonlinear effects decay faster and this term does not arise.

Refs. \cite{Hollands:2003ie,Ishibashi:2007kb,Tanabe:2009va,Tanabe:2011es} gave expressions for the rate of change of the Bondi energy at null infinity. For applications (e.g. higher-dimensional numerical relativity) it is convenient to have results that can be calculated easily and do not refer to a particular coordinate chart. This can be achieved by writing the result in terms of the asymptotic Weyl tensor components. We do this in section \ref{sec:bondiflux}. 

\section{Definitions of asymptotic flatness}

\label{sec:defs}

\subsection{Conformal definition for even $d$}

For even $d>4$, Ref. \cite{Hollands:2003ie} defined a spacetime $(M,g)$ to be asymptotically flat at null infinity as follows.  Given the (physical) metric $g$ and the Minkowski metric $\eta$, we would like to specify the precise rate at which $g$ approaches $\eta$ asymptotically.  We do this by conformally compactifying both $M$ and Minkowski spacetime so that ``infinity'' is now at a finite metric distance.  Thus, we obtain the ``unphysical'' spacetime $(\tilde{M},\tilde{g})$ and the ``background'' spacetime $(\bar{M},\bar{g})$, where the metrics $\tilde{g}$ and $\bar{g}$ are related to the respective physical and flat metrics via
\begin{equation}
\tilde{g}_{ab} = \Omega^2 g_{ab}, \qquad \bar{g}_{ab} = \Omega^2 \eta_{ab}
\end{equation}
with the conformal factor $\Omega^2$ satisfying the usual suitable properties.

Now, the spacetime is defined to be asymptotically flat at null infinity if 
\begin{gather}
 \tilde{g}_{ab}-\bar{g}_{ab}=\mc O (\Omega^{d/2-1}), \qquad \tilde{\epsilon}_{a_1\ldots a_d}-\bar{\epsilon}_{a_1\ldots a_d}=\mc O (\Omega^{d/2}),  \notag \\
(\tilde{g}^{ab}-\bar{g}^{ab})(\text{d}\Omega)_a=\mc O (\Omega^{d/2}), \qquad (\tilde{g}^{ab}-\bar{g}^{ab})(\text{d}\Omega)_a (\text{d}\Omega)_b=\mc O (\Omega^{d/2+1}),
\end{gather}
where $\tilde{g}^{ab}$ and $\bar{g}^{ab}$ are the inverse metrics of $\tilde{g}$ and $\bar{g}$, respectively and $\tilde{\epsilon}$ and $\bar{\epsilon}$ are the volume forms on $(\tilde{M},\tilde{g})$ and $(\bar{M},\bar{g})$, respectively.  Following Ref. \cite{geroch}, if $L_{ab\ldots c}$ is a tensor field on $\tilde{M}$ then the notation $L_{ab\ldots c}={\cal O}(\Omega^s)$ means that  $\Omega^{-s} L_{ab\ldots c}$ is smooth at future null infinity. 

\subsection{Definition using Bondi coordinates}

For general $d>4$, Ref. \cite{Tanabe:2011es} defined a spacetime to be asymptotically flat at future null infinity if, outside some cylindrical world tube, coordinates $(u,r,x^I)$ can be introduced following \cite{Sachs:1962wk} such that the metric takes the form
\be
\label{bondi}
 ds^2 = - Ae^B du^2 - 2e^B du dr + r^2 h_{IJ}(dx^I + C^I du)(dx^J + C^J du)
\ee
with
\be
\label{detcond}
 \det h_{IJ} = \det \omega_{IJ}
\ee
where $\omega_{IJ}(x)$ is the unit round metric on $S^{d-2}$. Surfaces of constant $u$ are null with topology $\mathbb{R} \times S^{d-2}$ where $x^I$ are coordinates on $S^{d-2}$ and $\mathbb{R}$ corresponds to the null geodesics generators of the surface. These generators are given by $u,x^I ={\rm constant}$ and $r$ is a (non-affine) parameter along the generators. $A$, $B$, $C^I$ and $h_{IJ}$ are functions of all of the coordinates. It is assumed that, at large $r$, they be expanded in inverse powers of $r$ (even $d$) or $\sqrt{r}$ (odd $d$) with\footnote{
Ref.  \cite{Tanabe:2011es} took $B={\cal O}(r^{-d})$, which was obtained by solving the vacuum Einstein equation. We have weakened this condition since we don't want to assume vacuum.}
\bea
\label{falloff}
 A &=& 1+{\cal O}(r^{-(d/2-1)}), \qquad B = {\cal O}(r^{-d/2}), \nonumber \\ C^I &=& {\cal O}(r^{-d/2}), \qquad h_{IJ} = \omega_{IJ} + {\cal O}(r^{-(d/2-1)}).
\eea
For odd $d$ it appears that an extra condition is required (discussed for $d=5$ in Ref. \cite{Tanabe:2009va}). One way of seeing this is to note that the results of Refs. \cite{Hollands:2003ie,Hollands:2004ac} suggest that, for linearized perturbations of Minkowski spacetime (arising from compactly supported initial data), the components of the metric perturbation each will be some half-integer power of $1/r$ times a {\it smooth} function of $1/r$. Hence each component will involve either integer powers of $1/r$ or half-odd-integer powers, but not both. Therefore, the presence of both integer and half-odd-integer powers in the expansions of individual metric components can be attributed to nonlinear effects. One would expect these only to affect terms beyond a certain order in the above expansions. If so, at low enough order, these expansions should contain only integer powers, or only half-odd-integer powers. This is indeed the case if one imposes the additional boundary condition that the expansion of $h_{IJ}$ in inverse powers of $\sqrt{r}$ contains no term of order $r^{-(d/2-1/2)}$ (see below).

\subsection{Equivalence of definitions for even $d$}

Starting from the Bondi definition, define $\Omega=1/r$. It is straightforward to show that this satisfies the conformal definition with (conformally flat) background metric
\be
 \bar{g} = -\Omega^2 du^2 + 2 du d\Omega + \omega_{IJ} dx^I dx^J.
\ee
Now consider a spacetime that is asymptotically flat according to the conformal definition. Write the flat metric in the form
\be
 \eta = -dU^2 - 2 dU dR + R^2  \omega_{IJ} (X) dX^I dX^J.
\ee
Now define $\Omega=1/R$. The background spacetime is 
\be
 \bar{g}= \Omega^2 \eta  = - \Omega^2 dU^2 + 2 dU d\Omega + \omega_{IJ} (X) dX^I dX^J.
\ee
${\cal I}^+$ is at $\Omega=0$ and $\Omega>0$ corresponds to the spacetime interior. 

In these coordinates, the definition of asymptotic flatness reduces to the following conditions on the unphysical spacetime
\bea
 \tilde{g}_{UU} &=& -\Omega^2 + {\cal O}(\Omega^{d/2+1}), \qquad \tilde{g}_{U\Omega} = 1 + {\cal O}(\Omega^{d/2}), \qquad \tilde{g}_{UI} = {\cal O}(\Omega^{d/2}) \nonumber \\
 \tilde{g}_{\Omega\Omega} &=& {\cal O}(\Omega^{d/2-1}), \qquad  \tilde{g}_{\Omega I } = {\cal O}(\Omega^{d/2-1}), \qquad \tilde{g}_{IJ} = \omega_{IJ} +  {\cal O}(\Omega^{d/2-1})
\eea
and
\be
 \det \tilde{g}_{IJ} = \det \omega_{IJ} + {\cal O}(\Omega^{d/2}).
\ee
Now convert to Gaussian null coordinates based on the null surface ${\cal I}^+$ in the unphysical spacetime as follows. Consider the (past-directed) null geodesic (of $\tilde{g}$) that passes through the point on ${\cal I}^+$ with coordinates $(u,0,x^I)$ and has tangent vector $\partial/\partial \Omega$ there. Let $\lambda$ denote the affine parameter along the geodesic. Since $\tilde{g}$ is required to be smooth near ${\cal I}^+$ it follows that the coordinates along the geodesic are smooth functions of $\lambda$ in a neighbourhood of $\lambda=0$. Expanding them in a Taylor series in $\lambda$ and substituting into the geodesic equations gives
\be
 U=u+ {\cal O}(\lambda^{d/2}), \qquad \Omega = \lambda +  {\cal O}(\lambda^{d/2+1}), \qquad X^I = x^I +  {\cal O}(\lambda^{d/2}).
\ee
We take $(u,\lambda,x^I)$ as new coordinates. In these coordinates, the unphysical metric is
\bea
 \tilde{g} &=& \left[ -\lambda^2 + {\cal O}(\lambda^{d/2+1}) \right] du^2 + 2 du d\lambda + {\cal O}(\lambda^{d/2}) du dx^I \nonumber \\ &+& \left[ \omega_{IJ}(x)+ {\cal O}(\lambda^{d/2-1}) \right]dx^I dx^J
\eea
where all components are smooth at $\lambda=0$ and
\be
 \det \tilde{g}_{IJ} = \det \omega_{IJ} + {\cal O}(\lambda^{d/2}).
\ee
We now replace $\lambda$ with a non-affine parameter $r$ defined by
\be
 r = \Omega^{-1} \left( \frac{ \det \tilde{g}_{IJ} }{\det \omega_{IJ}} \right)^{1/(2(d-2))} = \lambda^{-1}\left( 1 + {\cal O}(\lambda^{d/2}) \right)
\ee
so
\be
 \lambda = r^{-1}\left( 1 + {\cal O}(r^{-d/2}) \right)
\ee
and
\be
 \Omega^{-1} =  r \left( 1 + {\cal O}(r^{-d/2}) \right).
\ee
In coordinates $(u,r,x^I)$ the physical metric takes the Bondi form (\ref{bondi},\ref{detcond}) with metric coefficients that are smooth functions of $1/r$ respecting the fall-off conditions (\ref{falloff}).

\section{The Weyl tensor}

\label{sec:Weyl}

In this section, we determine the asymptotic fall off of Weyl tensor components for asymptotically flat spacetimes, as defined above. We will use Bondi coordinates since this allows us to treat even and odd $d$ simultaneously.  We perform our calculations using the higher dimensional Geroch-Held-Penrose (GHP) formalism of Ref. \cite{ghp} (see appendix \ref{app:ghp} for a review).

\subsection{Expansion of metric}

We begin with the metric written in Bondi coordinates (\ref{bondi},\ref{detcond}). From the definition of asymptotic flatness we have \cite{Tanabe:2011es}\footnote{Our notation differs slightly from that of Ref. \cite{Tanabe:2011es}, notably in the expansion coefficients of $B$.}
\begin{gather}
 h_{IJ}=\omega_{IJ}(x)+\sum_{k\geq 0} \frac{h^{(k+1)}_{IJ}(u,x)}{r^{d/2+k-1}}, \qquad  A =1+\sum_{k \geq 0}\frac{A^{(k+1)}(u,x)}{r^{d/2+k-1}}, \notag  \\
B =\sum_{k \geq 0} \frac{B^{(k+1)}(u,x)}{r^{d/2+k}}, \qquad
C^{I} = \sum_{k \geq 0} \frac{C^{(k+1)I}(u,x)}{r^{d/2+k}}, \label{asym:met}
\end{gather}
where in all of the summations $k \in \mathbb Z$ for even $d$ and $2k \in \mathbb Z$ for odd $d$.  Equation \eqref{detcond} implies that 
\be
 \omega^{IJ} h^{(k+1)}_{IJ}=0 \quad {\rm for}\quad  k<d/2-1
\ee
where $\omega^{IJ}$ is the inverse of $\omega_{IJ}$. In particular we have, for $d \ge 5$
\be
 \omega^{IJ} h^{(1)}_{IJ} = \omega^{IJ} h^{(3/2)}_{IJ} = \omega^{IJ} h^{(2)}_{IJ} = 0
\ee
and
\be
\omega^{IJ} h^{(5/2)}_{IJ} = 0 \quad (d>5), \qquad \omega^{IJ} h^{(5/2)}_{IJ} = \frac{1}{2} h^{(1)IJ} h^{(1)}_{IJ} \quad (d=5),
\ee 
where all indices on $h^{(k+1)}_{IJ}$ are raised using $\omega^{IJ}$. Some of the coefficients here have special significance. The Bondi mass is defined as \cite{Tanabe:2011es}
\be
\label{bondimass}
 M(u) = -\frac{d-2}{16\pi}  \int_{S^{d-2}} A^{(d/2-1)} d\omega
\ee
where the integral is taken over a sphere at null infinity. In vacuum it obeys the mass decrease law \cite{Tanabe:2011es}
\be
\label{bondiflux}
 \dot{M}(u) = -\frac{1}{32\pi} \int_{S^{d-2}} \dot{h}_{IJ}^{(1)}  \dot{h}^{(1)IJ }  d\omega.
\ee
This demonstrates that the quantity $h_{IJ}^{(1)}$ will be non-zero when gravitational radiation is present. Ref. \cite{Tanabe:2011es} showed that $h_{IJ}^{(1)}$ is not constrained by the asymptotic vacuum Einstein equation: it is a free function in the Bondi approach, just as in 4d \cite{bondi,Sachs:1962wk}. $\dot{h}^{(1)}_{IJ}$ corresponds to Bondi's ``news function".

\subsection{Null frame and connection components}

We choose a null frame $(\ell,n,\mb i)$ for the metric given by
\begin{gather}
 \ell =\mb 0= -\frac{\partial}{\partial r}, \quad n=\mb 1=e^{-B}\left( \frac{\partial}{\partial u} -\half A \frac{\partial}{\partial r} - C^I \frac{\partial}{\partial x^I} \right), \quad \mb i = e_{i}^{I} \frac{\partial}{\partial x^I}, \nn \\ \label{frame}
\ell^\flat=e^B du, \quad n^\flat = -(dr + \half A du), \quad \mb i ^{\flat} = e_{iI} (dx^I+C^Idu),
\end{gather}
where $e_i$ form a vielbein for the metric $h_{IJ}$ on $S^{d-2}$: $h_{IJ}=e_{i I} e_{jJ} \delta_{ij}$. We choose this vielbein by using the Gram-Schmidt algorithm starting from the basis $r^{-1} \hat{e}^I_i$ where the vectors $\hat{e}^I_i(x)$ form an orthonormal basis for the metric $\omega_{IJ}$ on $S^{d-2}$. This gives an expansion in inverse powers of $r$ (even $d$) or $\sqrt{r}$ (odd $d$):
\begin{equation}
 e_{i I} = r \left( \hat{e}_{iI} + \frac{e^{(1)}_{iI}}{r^{d/2-1}} \right) + \mc O (r^{-(d-5)/2}), \quad  e_{i}^{I} = r^{-1} \left( \hat{e}_{i}^{I} - \frac{e^{(1)I}_{i}}{r^{d/2-1}} \right) + \mc O (r^{-(d-1)/2}),
\end{equation}
where $2 \hat{e}_{i(I}e^{(1)}_{|j|J)} \delta_{ij}=h^{(1)}_{IJ}$ and $e^{(1)I}_{i} = \omega^{IJ} e^{(1)}_{iJ}$.

Using the definition of the connection components given in appendix \ref{app:ghp} and the null frame given in \eqref{frame}, we find that the GHP covariant connection components are
\begin{gather}
 \kappa_i =0, \quad \rho_{ij} = -\half e_i^I e_j^J \, \partial_{r}\left(r^2 h_{IJ}\right), \quad \rho=-(d-2)/r, \quad
\tau_i = -\half \left(e_i \cdot \partial B + e^{-B} e_{iI} \partial_{r} C^I \right) \nn \\
\kappa'_i =\half e^{-B} e_i \cdot \partial A, \quad \rho'_{ij} = - e^{-B} e_{(i|K|} e_{j)} \cdot \partial C^K +\half e_i^I e_j^J \, n \cdot \partial \left(r^2 h_{IJ}\right), \nn \\
\tau'_i = -\half \left(e_i \cdot \partial B - e^{-B} e_{iI} \partial_{r} C^I \right).
\end{gather}
The non-covariant coefficients are
\begin{gather}
 L_{10} = - \partial_{r}B, \quad L_{11} = \half e^{-B} \partial_{r}A, \quad L_{1i} = -\tau'_i \nn \\
\M i_{j0} = e^{I}_{[i} \partial_r e_{j]I}, \quad \M i_{j1} = e^{-B} e_{[i} \cdot \partial C^K e_{j]K} - e^{I}_{[i} n \cdot \partial e_{j]I} \nn \\
\M i_{jk} = r^2 e_i^I e_j^J e_k^K \partial_{[I} h_{J]K} - e^I_{[i} e_{|k|} \cdot \partial e_{j]I}.
\end{gather}
Using the asymptotic behaviour of the metric components given in \eqref{asym:met} gives
\begin{gather}
 \rho_{ij} = -\frac{\delta_{ij}}{r}+ \frac{\hat{e}^I_{i} \hat{e}^J_{j}}{4}  \left((d-2) \frac{ h^{(1)}_{IJ}}{r^{d/2}} +(d-1)\frac{ h^{(3/2)}_{IJ}}{r^{(d+1)/2}} \right) + \mc O(r^{-(d+2)/2}), \quad \rho=-(d-2)/r, \nn \\
\rho'_{ij} = - \frac{1}{2} \frac{\delta_{ij}}{r}+ \frac{\hat{e}^I_{i} \hat{e}^J_{j}}{2}  \left(\frac{ \dot{h}^{(1)}_{IJ}}{r^{d/2-1}} +\frac{\dot{h}^{(3/2)}_{IJ}}{r^{(d-1)/2}} \right) - \hat{e}_{(i}^I e_{j)}^{(1)J} \frac{\dot{h}^{(1)}_{IJ}}{r^{d-2}} + \mc O(r^{-d/2}), \nn \\
\rho' = -\frac{(d-2)}{2r} \left(\frac{1}{r}+\frac{A^{(1)}}{r^{d/2-1}}+\frac{A^{(3/2)}}{r^{(d-1)/2}}\right) + \mc O(r^{-d/2}), \quad \kappa'_i = \mc O(r^{-d/2}), \nn \\
\quad \tau_i = \frac{d}{4} \frac{\hat{e}\cdot C^{(1)}}{r^{d/2}} + \frac{d+1}{4} \frac{\hat{e}\cdot C^{(3/2)}}{r^{(d+1)/2}}+ \mc O(r^{-(d/2+1)}), \nn \\
\tau'_i = -\frac{d}{4} \frac{\hat{e}\cdot C^{(1)}}{r^{d/2}} - \frac{d+1}{4} \frac{\hat{e}\cdot C^{(3/2)}}{r^{(d+1)/2}} +\mc O(r^{-(d/2+1)}). \label{asym:spin1}
\end{gather}
where a dot denotes a partial derivative with respect to $u$. Also,
\begin{gather}
L_{10}=\frac{d}{2} \frac{B^{(1)}}{r^{d/2+1}} + \frac{(d+1)}{2} \frac{B^{(3/2)}}{r^{(d+3)/2}} +\mc O(r^{-(d/2+2)}), \nn \\
 L_{11}=-\frac{d-2}{4} \frac{A^{(1)}}{r^{d/2}} - \frac{d-1}{4} \frac{A^{(3/2)}}{r^{(d+1)/2}} +\mc O(r^{-(d/2+1)}), \quad L_{1i}=\mc O(r^{-d/2}), \nn \\
\M i_{j0} = \mc O(r^{-d/2}), \quad \M i_{j1} = - \frac{\hat{e}^I_{i} \dot{e}^{(1)}_{jI}}{r^{d/2-1}} + \mc O(r^{-(d-1)/2}), \nn \\
\eth_i=r^{-1} \hat{e}_i \cdot \nabla + \mc O (r^{-d/2}), \label{asym:spin2}
\end{gather}
where $\nabla_I$ denotes the covariant derivative induced by $\omega_{IJ}$. Of course, terms with half-odd-integer powers appear only for odd $d$.

\subsection{Parallely transported frame}

The null basis introduced above is convenient for calculations but it is not parallely transported along the geodesics. A parallely transported basis is one for which, in addition to the geodesic equation $\kappa_i=0$, we have $\tau'_i=L_{10}=\M i_{j0}=0$. Any such basis will be related to ours by a boost, spin and null rotation (see Appendix). Let $(\hat{\ell},\hat{n},\hat{m}_i)$ be such a basis. $\hat{\ell}$ must be parallel to $\ell$, with the coefficient fixed by requiring that $\hat{\ell}$ correspond to {\it affine} parameterization of the geodesics, ensuring $\hat{\kappa}_i = \hat{L}_{10} =0$. This gives $\hat{\ell}= e^{-B} \ell$, corresponding to a boost with parameter $e^{-B}$. $\tau'_i$ is invariant under a boost and transforms covariantly under a spin. But under a null rotation with parameters $z_i$ it transforms inhomogeneously \cite{ghp} so $z_i$ is determined by $\hat{\tau}'_i=0$. This gives
\be
 z_i = c_i + {\cal O}(r^{-(d/2-1)} )
\ee
where the parameters $c_i$ are independent of $r$. Finally, $\M i_{j0}$ transforms homogeneously under a boost and trivially under a null rotation but inhomogeneously under a spin. Requiring $\stackrel{i}{\hat{M}}_{j0}=0$ determines the spin matrix to be
\be
\label{spinparallel}
 X_{ij} = O_{ij} + {\cal O}(r^{-(d/2-1)})
\ee
where $O_{ij}$ is a $r$-independent orthogonal matrix. 

Our strategy will be determine curvature components in the basis defined previously and then transform our results to a parallely transported frame by first performing a boost with parameter $e^{-B} = 1 + {\cal O}(r^{-d/2})$, then a null rotation with parameters $z_i$ and finally a spin with parameters $X_{ij}$ as given above.

\subsection{Calculation of curvature components}

In the GHP formalism the Weyl tensor components are denoted
\begin{gather}
 \Omega_{ij} = C_{0i0j}, \qquad \Psi_{ijk} = C_{0ijk}, \qquad \Psi_{i} = C_{010i} = \Psi_{jij}, \notag \\
\Phi_{ijkl} = C_{ijkl}, \qquad \Phi_{ij} = C_{0i1j}, \qquad \Phi=\Phi_{ii}=C_{0101}, \notag \\
(2 \Phis_{ij}= 2 \Phi_{(ij)} = - \Phi_{ikjk},  \quad 2 \Phia_{ij}= 2 \Phi_{[ij]} = C_{01ij}), \notag \\
\Omega'_{ij} = C_{1i1j}, \qquad \Psi'_{ijk} = C_{1ijk}, \qquad \Psi'_{i} = C_{101i} = \Psi'_{jij} \label{weylcomps}
\end{gather}
and the Ricci tensor components are
\begin{gather}
\omega= R_{00}, \quad \psi_{i} = R_{0i}, \quad \phi_{ij} = R_{ij}, \quad \phi=R_{01}, \quad \psi'_{i} = R_{1i}, \quad \omega' = R_{11}.
\end{gather}
The Newman-Penrose equations (see Appendix) are used to determine all of these quantities except for those of boost weight zero (i.e. those written with the letters $\Phi$ or $\phi$). To determine the latter we used the Bianchi equation (B3) from the Appendix.\footnote{
This involves an integration with respect to $r$, introducing a homogeneous term decaying as $1/r$ into the boost weight zero quantities. This is not compatible with asymptotic flatness so the coefficient of this term must vanish. This could be shown e.g. by using the commutator (C3) of Ref. \cite{ghp}.}

\subsection{Results: even $d$}

In our basis (\ref{frame}), we find that the Ricci tensor components are smooth functions of $1/r$ with
\begin{gather}
 \omega=\mc O(r^{-(d/2+2)}), \qquad \psi_i=\mc O(r^{-(d/2+1)}), \qquad \phi_{ij}=\mc O(r^{-(d/2+1)}),  \nn \\
\phi=\mc O(r^{-(d/2+1)}), \qquad \psi'_i = \mc O(r^{-d/2}), \qquad \omega' = \mc O(r^{-d/2}).
  \label{riccifalloff}
\end{gather} 
The Weyl tensor components are smooth functions of $1/r$ with
\begin{gather}
 \Omega_{ij} = -\frac{(d-2)(d-4)}{8} \frac{\hat{e}^I_{i} \hat{e}^J_{j} h^{(1)}_{IJ}}{r^{d/2+1}} + \mc O(r^{-(d/2+2)}), \nn \\
 \Psi_{ijk} = \mc O(r^{-(d/2+1)}), \qquad  \Psi_i=\mc O(r^{-(d/2+1)}), \qquad \Phi^A_{ij}=\mc O(r^{-(d/2+1)}),\qquad \Phi=\mc O(r^{-(d/2+1)}), \nn \\
\Phi^S_{ij} = -\frac{(d-4)}{4} \frac{\hat{e}^I_{i} \hat{e}^J_{j} \dot{h}^{(1)}_{IJ}}{r^{d/2}} + \mc O(r^{-(d/2+1)}), 
\qquad \Phi_{ijkl} = (\hat{e}_i^I \hat{e}^J_{[k}\delta_{l]j} - \hat{e}_j^I \hat{e}^J_{[k}\delta_{l]i}) \frac{\dot{h}^{(1)}_{IJ}}{r^{d/2}} + \mc O(r^{-(d/2+1)}), \nn \\
\Psi'_{ijk}=\mc O(r^{-d/2}), \qquad \Psi'_i=\mc O(r^{-d/2}), \nn \\
\Omega'_{ij} = - \frac{1}{2} \frac{\hat{e}_i^I \hat{e}_j^J \ddot{h}^{(1)}_{IJ}}{r^{d/2-1}} + \mc O(r^{-d/2}). \label{Weyl:even}
\end{gather}
Recall that $h^{(1)}_{IJ}$ is non-zero in any spacetime containing outgoing gravitational radiation, and it is not determined by the asymptotic Einstein equation. 

Now we transform to a parallely transported frame as determined above. The boost and spin are easy to deal with since the curvature components transform covariantly with respect to these. Formulae for the transformation under a null rotation are given in the Appendix. Using these results, we see that the transformation to a parallely transported frame does not change any of these results (aside from acting with the rotation matrix $O_{ij}$ on the indices $i,j,k$ etc). 

Finally we have to convert from our parameter $r$ to an affine parameter along the geodesics. Denote the latter by $\lambda$. Then (up to the freedom to multiply by a quantity independent of $r$)
\be
 \lambda = \int e^B dr = r + c + {\cal O}(r^{-(d/2-1)})
\ee
where $c$ is independent of $r$. Inverting gives
\be
\label{affine}
 r=\lambda - c + {\cal O}(\lambda^{-(d/2-1)}).
\ee
If we substitute this into the above expressions for the Weyl components then they become smooth functions of $1/\lambda$ with leading order behaviour given by replacing $r$ with $\lambda$ in these expressions. Hence the leading order term in the Weyl tensor is of order $\lambda^{-(d/2-1)}$ and the only non-vanishing components at this order are $\Omega'_{ij}$, which (from (\ref{Weyl:even})) is generically non-zero. But this is precisely the definition of a type N Weyl tensor with $\ell$ (the tangent to the geodesics) a multiple Weyl Aligned Null Direction (WAND) \cite{Coley:2004jv}.

The next non-vanishing terms in the Weyl arise at order $\lambda^{-d/2}$. Such terms can arise from $\Omega'_{ij}$, $\Psi'_{ijk}$, $\Psi'_i$, $\Phi_{ijkl}$ and $\Phi^S_{ij}$. So, at this order, we have $\Omega_{ij} = \Psi_{ijk} = \Psi_i=0$ and hence the Weyl tensor is type II with multiple WAND $\ell$. It cannot be type III because (\ref{Weyl:even}) shows that $\Phi^S_{ij}$ is generically non-vanishing. However, it is not the most general possible type II Weyl tensor because (\ref{Weyl:even}) shows that it has vanishing $\Phi$ and $\Phi^A_{ij}$.

After this, we have terms of order $\lambda^{-(d/2+1)}$. At this order, any of the Weyl components can be non-zero. In particular, the above expression shows that $\Omega_{ij}$ is generically non-zero, which implies that the Weyl tensor at this order is type G (i.e. $\ell$ is not a WAND).

In summary, for even $d>4$, we have demonstrated that, in a spacetime satisfying the definition of asymptotic flatness at null infinity of Refs. \cite{Hollands:2003ie,Tanabe:2011es}, the Weyl tensor exhibits the peeling behaviour described around equation (\ref{peeling}), with the type II part obeying the additional conditions $\Phi=\Phi^A_{ij}=0$.

When $d=4$, our results for $\Omega'_{ij}$, $\Psi'_{ijk}$ and $\Psi'_i$ are consistent with the 4d peeling property. The boost-weight zero terms also are consistent: in 4d, all such terms are determined by $\Phi$ and $\Phi^A_{ij}$, which vanish at order $\lambda^{-d/2}=\lambda^{-2}$. Hence at order $\lambda^{-2}$ we have a Weyl tensor of type III instead of type II. More explicitly,  in 4d, $\Phi_{ijkl}$ is determined by its trace $\Phi^S_{ij}$. But the first term in the expansion of $\Phi^S_{ij}$ in (\ref{Weyl:even}) comes with a coefficient of $d-4$. Similar results hold for the other Weyl components (e.g. the above expression for $\Omega_{ij}$ has a factor $d-4$). This is why peeling is qualitatively different when $d=4$.

\subsection{Results: odd $d$}

As discussed above, for odd $d$ there is an additional condition in the definition of asymptotic flatness at future null infinity, that the term of order $r^{-(d/2-1/2)}$ in the expansion of $h_{IJ}$ should be absent, i.e.,
\be
\label{h32}
 h^{(3/2)}_{IJ} = 0.
\ee
With this condition, we find that the results (\ref{riccifalloff},\ref{Weyl:even}) for the Ricci and Weyl components are valid also for odd $d$, with the understanding that these formulae now refer to expansions in inverse powers of $\sqrt{r}$. (Without (\ref{h32}), there would be e.g. a term of order $r^{-(d/2-1/2)}$ in the expansion of $\Omega'_{ij}$.) Following the same steps as for even $d$, converting to a parallely transported frame and affine parameterization, we find, just as before, that the leading components of the Weyl tensor arise at order $\lambda^{-(d/2-1)}$ and this term is type N as before. There are no terms at order $\lambda^{-(d/2-1/2)}$ so the next term is at order $\lambda^{-d/2}$ which is type II with $\Phi^A_{ij} = \Phi=0$, again as for even $d$. 

A difference between even and odd $d$ arises at next order: for odd $d$ there is the possibility of terms of order $r^{-(d/2+1/2)}$. For example, we find that
\begin{align}
 \Omega'_{ij} + \frac{\omega'}{d-2} \delta_{ij}=& - \frac{1}{2} r^{-(d/2-1)} \hat{e}_i^I \hat{e}_j^J \ddot{h}^{(1)}_{IJ}+ r^{-d/2} Y_{ij} +\frac{1}{4} r^{-(d-2)}
   \left(4 \hat{e}^{I}_{(i} e^{(1)J}_{j)} \ddot{h}^{(1)}_{IJ}+ \hat{e}_i^{I} \hat{e}_j^{J} \omega^{KL} \dot{h}^{(1)}_{IK} \dot{h}^{(1)}_{JL}\right) \nn \\
& -\frac{1}{2} r^{-(d/2+1/2)} \left( \hat{e}_i^I \hat{e}_j^J \ddot{h}^{(5/2)}_{IJ}  - \dot{A}^{(3/2)} \delta_{ij} -2 \hat{e}_{i|K|} \hat{e}_{j)} \cdot \nabla \dot{C}^{(3/2)K} \right)   +{\cal O}(r^{-(d/2+1)}), \label{Omegaprime}
\end{align}
where $Y_{ij}$ is a quantity whose explicit form we will not need. The Weyl components $\Omega'_{ij}$ are obtained by taking the traceless part of this equation and the Ricci component $\omega'$ by taking the trace. We have retained a term of order $r^{-(d-2)}$ because $r^{-(d-2)} = r^{-(d/2+1/2)}$ if $d=5$. 

The only other Weyl components containing terms of order $r^{-(d/2+1/2)}$ are $\Psi'_{ijk}$ and $\Psi'_i$, which can be obtained from
\bea
 \Psi'_{ijk} + \frac{2}{d-2} \psi'_{[j} \delta_{k]i} &=&  r^{-d/2} \hat{e}_i^I \hat{e}^J_{[j} \hat{e}_{k]} \cdot \nabla \dot{h}^{(1)}_{IJ}+ \mc O(r^{-(d/2+1)}),\nn \\  \Psi'_i - \frac{1}{d-2} \psi'_i &=& \frac{d}{4} r^{-d/2}  \hat{e}_i \cdot  \dot{C}^{(1)} + \frac{(d+1)}{4}  r^{-(d/2+1/2)} \hat{e}_i \cdot \dot{C}^{(3/2)} + \mc O(r^{-(d/2+1)}) \label{Psiprime}
\eea
where the Weyl and Ricci components can be disentangled by taking a trace of the first equation and combining with the second equation.  

We will now argue that terms of order $r^{-(d/2+1/2)}$ can be eliminated for $d>5$ by exploiting the Einstein equation (which we did not use for even $d$). We will assume that the Ricci tensor (and hence the energy-momentum tensor) decays faster near infinity than the rate which is given by asymptotic flatness alone (equation (\ref{riccifalloff})). The rate that we require is faster by a factor $1/r$:\footnote{For $d>5$, the constraints on $\omega$, $\psi_i$ and $\phi$, necessarily 
imply the constraints on $\psi'_i$ and $\omega'$. For $d=5$, the constraint on $\psi_i$ implies the constraint on $\psi'_i$.}
\begin{gather}
 \omega=\mc O(r^{-(d/2+3)}), \qquad \psi_i=\mc O(r^{-(d/2+2)}), \qquad \phi_{ij}=\mc O(r^{-(d/2+2)}),  \nn \\
\phi=\mc O(r^{-(d/2+2)}), \qquad \psi'_i = \mc O(r^{-(d/2+1)}), \qquad \omega' = \mc O(r^{-(d/2+1)}).
  \label{riccicons}
\end{gather} 
Imposing these conditions implies that the first few coefficients in the expansions of the metric components must satisfy the same equations as in a {\it vacuum} spacetime, as determined in Ref. \cite{Tanabe:2011es}:
\begin{gather}
 B^{(1)}=0, \qquad  A^{(1)}=- \frac{2}{d-2} \nabla \cdot C^{(1)}=-\frac{4}{d(d-2)} \nabla^I \nabla^{J} h^{(1)}_{IJ},
\qquad C^{(1)I} = \frac{2}{d} \nabla_{J}h^{(1)IJ}, \nn \\
 \begin{cases} \dot{A}^{(3/2)}=\frac{1}{6}\dot{h}^{(1) IJ} \dot{h}^{(1)}_{IJ} & d=5 \\A^{(3/2)}= 0 & d>5   \end{cases}, \qquad
B^{(3/2)}= \begin{cases} -\frac{1}{16}h^{(1) IJ} h^{(1)}_{IJ} & d=5 \\ 0 & d>5   \end{cases}, \nn \\
C^{(3/2)I}=0, \qquad \dot{h}^{(5/2)}_{IJ}= \begin{cases} \omega^{KL} h^{(1)}_{K(I} \dot{h}^{(1)}_{J)L} & d=5 \\ 0 & d>5   \end{cases}
\end{gather}
and an equation relating $\dot{h}^{(2)}_{IJ}$ to $A^{(1)}$, $C^{(1)I}$ and $h^{(1)}_{IJ}$. Note that the asymptotic Einstein equation implies no restriction on $h^{(1)}_{IJ}$. Recall that for $d=5$, $A^{(3/2)}$ determines the Bondi mass via (\ref{bondimass}).\footnote{\label{strongerasymp}
Note also that $\dot{h}_{IJ}^{(5/2)}=0$ for $d>5$ implies that one can impose the additional boundary condition $h^{(5/2)}_{IJ}=0$ for $d>5$. Ref. \cite{Tanabe:2011es} examined the vacuum Einstein equations to higher order and the results suggest that the definition of asymptotic flatness for odd $d$ should be augmented with the condition $h^{(k+1)}_{IJ}=0$ for $k=1/2, 3/2, \ldots , d/2-2$ although we will not assume any more than (\ref{h32}).}

Using these results, we see that the term of order $r^{-(d/2+1/2)}$ in (\ref{Psiprime}) is absent and hence such terms do not appear in $\Psi'_{ijk}$ and $\Psi'_i$. However, terms of this order are absent from (\ref{Omegaprime}) if, and only if, $d>5$. Hence, for $d>5$, such terms are absent from $\Omega'_{ij}$.  Transforming to a parallely transported frame and affine parameterization, similar arguments to those used for the even $d$ case establish the peeling result given in equation \eqref{peeling} for odd $d>5$. As for even $d$, the type II term obeys the additional restrictions $\Phia_{ij}=\Phi=0$.

Finally we must discuss the $d=5$ case. For $d=5$, terms of order $r^{-(d/2+1/2)}=r^{-3}$ do {\it not} drop out of $\Omega'_{ij}$:
\begin{align}
\Omega'_{ij}= &- \frac{1}{2} \frac{\hat{e}_i^I \hat{e}_j^J \ddot{h}^{(1)}_{IJ}}{r^{3/2}} + \frac{Y_{ij}}{r^{5/2}} \nn \\
&-\frac{1}{2r^3} \left\lbrace \hat{e}_i^I \hat{e}_j^J \left(\omega^{KL} h^{(1)}_{IK} \ddot{h}^{(1)}_{JL} +\textstyle{\frac{1}{2}} \omega^{KL} \dot{h}^{(1)}_{IK} \dot{h}^{(1)}_{JL} -\textstyle{\frac{1}{6}} \omega_{IJ} \dot{h}^{(1)KL} \dot{h}^{(1)}_{KL} \right) - 2 \hat{e}^{I}_{(i} e^{(1)J}_{j)} \ddot{h}^{(1)}_{IJ} \right\rbrace +{\cal O}(r^{-7/2}).
\end{align}
Note that the coefficient of $r^{-3}$ is quadratic in $h^{(1)}_{IJ}$ and its time derivatives and hence generically it is non-zero if gravitational radiation is present.

Now we must transform to a parallelly transported frame. As for $d>5$, the boost and null rotation do not change our results. But note that the spin matrix (\ref{spinparallel}) involves a term of order $r^{-3/2}$. Hence when the spin acts on $\Omega'_{ij}$ this term will combine with the leading term in $\Omega'_{ij}$ to produce a new term of order $r^{-3}$ in $\hat{\Omega}'_{ij}$. Could this new term cancel the terms already present? Generically no: the new term will involve $\ddot{h}^{(1)}_{IJ}$ whereas some of the terms already present involve only first derivatives of $h^{(1)}_{IJ}$. Since $\dot{h}^{(1)}_{IJ}$ is a free function in the Bondi approach, these terms will not cancel in general. For example, one could choose $\ddot{h}^{(1)}_{IJ}$ to be zero somewhere, with $\dot{h}^{(1)}_{IJ}$ non-zero. 

The last step is to convert to affine parameterization using (\ref{affine}), which does not change anything. We conclude that for $d=5$, the Weyl tensor satisfies the peeling property (\ref{peeling5d}) described in the introduction. Again the type II term obeys the additional restrictions $\Phia_{ij}=\Phi=0$.

\section{Bondi flux}

\label{sec:bondiflux}

In 4d, the rate of decrease of the Bondi energy at future null infinity is given in terms of the Newman-Penrose Weyl scalar $\Psi_4$ as
\be
 \dot{M}(u)  =- \lim_{r \to \infty} \frac{r^2}{4\pi} \int_{S^2} \left| \int_{-\infty}^u \Psi_4(\hat{u},r,x) d\hat{u} \right|^2 d\omega
\ee
where $d\omega$ is the volume element on a unit $S^2$. In $d>4$ dimensions, the rate of decrease of the Bondi energy at future null infinity is given by (\ref{bondiflux}) \cite{Tanabe:2011es}. We can rewrite this in terms of $\Omega'_{ij}$ (the analogue of $\Psi_4$) as follows. Assume that the Bondi flux vanishes in the far past, i.e. $\dot{h}^{(1)}_{IJ} \rightarrow 0$ as $u \rightarrow -\infty$. Then from (\ref{Weyl:even}) (which holds for even or odd $d$) we have
\be
 \hat{e}_i^I \hat{e}_j^J \dot{h}^{(1)}_{IJ}(u,x) = -2 \lim_{r\rightarrow \infty} r^{d/2-1} \int_{-\infty}^u \Omega'_{ij} (\hat{u}, r, x) \, d\hat{u}
\ee
and hence
\be
 \dot{M}(u) = - \lim_{r\rightarrow \infty} \frac{r^{d-2}}{8 \pi} \int_{S^{d-2}} \left(  \int_{-\infty}^u \Omega'_{ij} (\hat{u}, r, x) \, d\hat{u} \right)^2 \, d\omega 
\ee
where $d\omega$ is the volume element on a unit $S^{d-2}$ and $(Y_{ij})^2 \equiv Y_{ij} Y_{ij}$. In practice, the RHS is computed by choosing coordinates so that the asymptotic metric takes the form
\be
 ds^2 \sim -du^2 -2du dr + r^2 d\omega^2.
\ee
One then chooses a null vector field $n$ that approaches $\pm (\partial/\partial u - \half \partial/\partial r)$ asymptotically (the sign does not matter here) and a set of orthonormal spacelike vectors $m_{(i)}$ ($i=2, \ldots d-1$) such that $n \cdot m_{(i)} = 0$. Then $\Omega'_{ij} = C_{abcd} n^a m_{(i)}^b n^c m_{(j)}^d$.

\section{Discussion}

We derived our result using Bondi coordinates since this allows us to treat even and odd $d$ together for much of the analysis. However, for even $d$ it would be more elegant to derive the peeling property using the conformal approach. It would be nice to see this worked out.

For odd $d>5$, our result (\ref{peeling}) involves only inverse half-odd-integer powers of $\lambda$. Inverse integer powers will appear if one continues to higher orders in the expansion. It would be interesting to know at what order inverse integer powers first appear. If one strengthens the definition of asymptotic flatness as suggested in footnote \ref{strongerasymp} then it seems likely that the first such terms will appear at order $\lambda^{-(d-2)}$, in agreement with our result for $d=5$.

Ref. \cite{Ortaggio:2009zt} studied asymptotically flat solutions in $d>4$ dimensions that are {\it algebraically special}. It was found that the latter condition is incompatible with gravitational radiation (in contrast with the $d=4$ case). We can see a similar result here: if $\ell$ is a WAND then $\Omega_{ij}$ must vanish. For $d>4$, (\ref{Weyl:even}) then requires $h^{(1)}_{IJ}=0$, which implies vanishing Bondi energy flux, i.e., no gravitational radiation.

\bigskip

\begin{center} {\bf Acknowledgments} \end{center}

\noindent We are grateful to Mihalis Dafermos for useful discussions and to Vojtech Pravda and Alena Pravdova for comments on a draft. MG is supported by St John's College, Cambridge. HSR is a Royal Society University Research Fellow. HSR acknowledges support from the European Research Council under grant agreement ERC-2011-StG 279363-HiDGR. 

\appendix

\section{Higher dimensional GHP formalism} \label{app:ghp}

In this appendix, we review the higher dimensional GHP formalism of \cite{ghp}.  Given a background solution, we choose a null frame $(\ell, n ,\mb i)$ such that in this frame, the metric takes the form
\be \label{metric}
g_{\mu \nu}=2 \ell_{(\mu} n_{\nu)}+ \mb i_{\mu} \mb i_{\nu}.
\ee

In the GHP formalism, one breaks complete covariance by singling out two null directions ($\ell$ and $n$) at each point, but preserves covariance in the remaining directions.  This is in contrast to the NP formalism where none of the covariance is preserved.

At any point, the Lorentz group is generated by
\begin{itemize}
 \item boosts ($\mu$ a real function):
\be
\ell \rightarrow \mu \, \ell, \quad n \rightarrow \mu^{-1} n, \quad \mb i \rightarrow \mb i, \label{app:boost}
\ee
 \item spins ($X_{ij} \in SO(d-2)$): 
\be
\ell \rightarrow \ell, \quad n \rightarrow n, \quad \mb i \rightarrow X_{i j} \mb j, \label{app:spin}
\ee
\item null rotations about $\ell$ ($z_i$ $d-2$ real functions):
\be
\ell \rightarrow \ell, \quad n \rightarrow n + z_i \mb i - \half z^2 \ell, \quad \mb i \rightarrow \mb i - z_i \ell, \label{app:rotl}
\ee
\end{itemize}
where $\lambda \neq 0$ and $X_{ij}$ is some position-dependent orthogonal matrix.

We would like to keep the subgroup that preserves the null directions, i.e. the subgroup given by boosts and spatial rotations, or spins.  Thus, we would like to work with objects that transform covariantly under this subgroup.

A {\it GHP scalar of boost weight $b$ and spin $s$} is a scalar $\eta_{i_1 \ldots i_{s}}$ that transforms covariantly as
\be
 \eta_{i_1 \ldots i_{s}} \rightarrow \mu^{b} \eta_{i_1 \ldots i_{s}}
\ee
under boosts and 
\be
 \eta_{i_1 \ldots i_{s}} \rightarrow X_{i_1 j_1} \cdots X_{i_s j_s} \eta_{j_1 \ldots j_{s}}
\ee
under spins. Evidently, the product of two GHP scalars of boost weights $b_1$ and $b_2$ and spins $s_1$ and $s_2$, respectively, gives a GHP scalar of boost weight $b_1+b_2$ and spin $s_1+s_2$.

Denote the covariant derivatives of the basis vectors as
\be
L_{\mu \nu} = \nabla_{\nu} \ell_\mu, \quad N_{\mu \nu} = \nabla_{\nu} n_\mu, \quad \M i_{\mu \nu} = \nabla_{\nu} \mb i_\mu,
\ee
Not all the scalars formed from the projection of these objects into the basis are GHP scalars.  Those that are GHP scalars are listed in table \ref{tab:npsca} \cite{ghp}.

\begin{table}[ht]
\caption{GHP scalars constructed from covariant derivatives of the basis vectors.} \centering
\vspace{1.5mm}
\begin{tabular}{c c c c l}
\hline
Spin coefficient & GHP notation & Boost weight $b$ & Spin $s$ & Interpretation\\ [1mm]
\hline 
$L_{ij}$  & $\rho_{ij}$  & 1  & 2 & expansion, shear and twist of $\lb$\\[1mm]
     $L_{ii}$  & $\rho=\rho_{ii}$  & 1  & 0 & expansion of $\lb$\\[1mm]
    $L_{i0}$  & $\kap_{i}$   & 2  & 1 & non-geodesity of $\lb$\\[1mm]
    $L_{i1}$  & $\tau_{i}$   & 0  & 1 & transport of $\lb$ along $n$\\[1mm]
    $N_{ij}$  & $\rho'_{ij}$ & -1 & 2 & expansion, shear and twist of $n$\\[1mm]
     $N_{ii}$  & $\rho'=\rho'_{ii}$ & -1 & 0 & expansion of $n$\\[1mm]
     $N_{i1}$  & $\kap'_{i}$  & -2 & 1 & non-geodesity of $n$\\[1mm]
    $N_{i0}$  & $\tau'_{i}$  & 0  & 1 & transport of $n$ along $l$\\[1mm]\hline
\end{tabular}
\label{tab:npsca}
\end{table}

Notice that we have used a prime operation, which interchanges the null basis vectors
\be
' \ : \  \ell \leftrightarrow n.
\ee
For a GHP scalar $\eta_{i_1 \ldots i_{s}}$ of boost weight $b$ and spin $s$, we define its GHP covariant derivatives to be\footnote{Symbols $\tho$ and $\eth$, pronounced ``thorn'' and ``eth'', respectively are old Germanic letters that have been retained in the Icelandic alphabet.}
 \begin{eqnarray} \label{ghpdertho}
    \tho T_{i_1 i_2...i_s} &\equiv & \ell \cdot \partial T_{i_1 i_2...i_s} - b L_{10} T_{i_1 i_2...i_s} 
                                     + \sum_{r=1}^s \M{k}_{i_r 0} T_{i_1...i_{r-1} k i_{r+1}...i_s},\\
    \tho' T_{i_1 i_2...i_s} &\equiv & n \cdot \partial T_{i_1 i_2...i_s} - b L_{11} T_{i_1 i_2...i_s} 
                                     + \sum_{r=1}^s \M{k}_{i_r 1} T_{i_1...i_{r-1} k i_{r+1}...i_s}, \label{ghpdertho'} \\ \label{ghpderm}
    \eth_i T_{j_1 j_2...j_s} &\equiv & \mb i \cdot \partial T_{j_1 j_2...j_s} - b L_{1i} T_{j_1 j_2...j_s} 
                                     + \sum_{r=1}^s \M{k}_{j_r i} T_{j_1...j_{r-1} k j_{r+1}...j_s}.
  \end{eqnarray}

GHP versions of the "Newman-Penrose" and Bianchi equations are given in Ref. \cite{ghp}. We will need the NP equations:
\newcounter{oldeq}
\setcounter{oldeq}{\value{equation}}
\renewcommand{\theequation}{NP\arabic{equation}}
\setcounter{equation}{0}
\begin{eqnarray}\label{NP1}
  \tho \rho_{ij} - \eth_j \kap_i &=& - \rho_{ik} \rho_{kj} -\kap_i \tau'_j - \tau_i \kap_j - \Om_{ij} - \textstyle{\frac{1}{d-2}} \omega \delta_{ij} ,\\[3mm]
  \tho \tau_i - \tho' \kap_i &=& \rho_{ij}(-\tau_j + \tau'_j) - \Psi_i + \textstyle{\frac{1}{d-2}} \psi_{i},\label{NP2}\\[3mm]
  2\eth_{[j|} \rho_{i|k]}     &=& 2\tau_i \rho_{[jk]} + 2\kap_i \rho'_{[jk]} - \Psi_{ijk} - \textstyle{\frac{1}{d-2}} \psi_{[j} \delta_{k]i},\label{R:ethrho}\label{NP3}\\[3mm]
  \tho' \rho_{ij} - \eth_j \tau_i &=& - \tau_i \tau_j - \kap_i \kap'_j - \rho_{ik}\rho'_{kj}-\Phi_{ij}
                                       -\textstyle{\frac{1}{d-2}} (\phi_{ij}+\phi \delta_{ij}) + \textstyle{\frac{\phi_{kk} + 2\phi}{(d-1)(d-2)}} \delta_{ij} .\label{NP4}
\end{eqnarray}
Another four equations can be obtained by taking the prime $'$ of these four (i.e.\ by exchanging the vectors $\lb$ and $\nb$). We also need the following components of the Bianchi identity:
\renewcommand{\theequation}{A.\arabic{equation}}
\setcounter{equation}{\value{oldeq}}
\setcounter{oldeq}{\value{equation}}
\renewcommand{\theequation}{B\arabic{equation}}
\setcounter{equation}{2}
\begin{eqnarray}
  -\tho \tilde{\Phi}_{ijkl} + 2 \eth_{[k}\tilde{\Psi}_{l]ij}
                 &=& - 2 \tilde{\Psi}'_{[i|kl} \kap_{|j]} - 2 \tilde{\Psi}'_{[k|ij}\kap_{|l]}\nn\\
                 &&  + 4{\tilde{\Phi}^{\text{A}}}_{ij} \rho_{[kl]} -2\tilde{\Phi}_{[k|i}\rho_{j|l]} 
                     + 2\tilde{\Phi}_{[k|j}\rho_{i|l]} + 2 \tilde{\Phi}_{ij[k|m}\rho_{m|l]}\nn\\
                 &&  -2\tilde{\Psi}_{[i|kl}\tau'_{|j]} - 2\tilde{\Psi}_{[k|ij} \tau'_{|l]}
                     - 2\tilde{\Omega}_{i[k|} \rho'_{j|l]} + 2\tilde{\Omega}_{j[k} \rho'_{i|l]},
                     \label{B3}
\end{eqnarray}
\renewcommand{\theequation}{\arabic{equation}}
\setcounter{equation}{\value{oldeq}}
where the tilde notation indicates components of the Riemann tensor analogously defined to those of the components of the Weyl tensor given in \eqref{weylcomps}.  The relation of these components to the components of the Weyl and Ricci tensors is given by the definition of the Weyl tensor as the trace-free part of the Riemann tensor
\begin{gather}
 \Omega_{ij}= \tilde{\Omega}_{ij} -\frac{\omega}{d-2} \delta_{ij}, \quad \Psi_{ijk}= \tilde{\Psi}_{ijk} - \frac{2}{d-2} \psi_{[j}\delta_{k]i}, \quad \Psi_{i}=\tilde{\Psi}_i +\frac{1}{d-2}\psi_i , \nn \\
\Phi_{ijkl}=\tilde{\Phi}_{ijkl} - \frac{2}{d-2}\left( \phi_{i[k}\delta_{l]j} - \phi_{j[k}\delta_{l]i} \right) + \frac{2}{(d-1)(d-2)} (2\phi+\phi_{mm}) \delta_{i[k}\delta_{l]j} , \nn \\
\Phi_{ij}= \tilde{\Phi}_{ij} - \frac{1}{d-2}\left(\phi \delta_{ij}+ \phi_{ij}\right) + \frac{(2\phi+\phi_{mm})}{(d-1)(d-2)}  \delta_{ij}, \quad \Phi=\tilde{\Phi} +\frac{2\phi}{d-1}-\frac{\phi_{ii}}{(d-1)(d-2)}, \nn \\
\Omega'_{ij}= \tilde{\Omega}'_{ij} -\frac{\omega'}{d-2} \delta_{ij}, \quad \Psi'_{ijk}= \tilde{\Psi}'_{ijk} - \frac{2}{d-2} \psi'_{[j}\delta_{k]i}, \quad \Psi'_{i}=\tilde{\Psi}'_i +\frac{1}{d-2}\psi'_i.
\end{gather}

\subsection*{Null rotations}
Under a null rotation about $\ell$ of the form given by equation \eqref{app:rotl} the Weyl tensor components transform as:
\begin{eqnarray}
  \Om_{ij}  &\mapsto &\Om_{ij},\\
  \Ps_{i}   &\mapsto &\Ps_{i}+\Om_{ij}z_j,\\
  \Ps_{ijk} &\mapsto &\Ps_{ijk}+2\Om_{i[j}z_{k]},\\
  \Phi      &\mapsto &\Phi + 2z_i\Ps_i + z_i \Om_{ij}z_j,\\
  \Phi_{ij}  &\mapsto& \Phi_{ij}+z_j\Ps_i+z_k\Ps_{ikj}+Z_{jk}\Om_{ik} ,\ \\
  \Phi_{ijkl}&\mapsto& \Phi_{ijkl} - 2z_{[k}\Ps_{l]ij} - 2z_{[i}\Ps_{j]kl}-2z_jz_{[k}\Om_{l]i}+2z_iz_{[k}\Om_{l]j},\\
  \Ps'_i      &\mapsto& \Ps'_i - z_i\Phi + 3\Phia_{ij}z_j - \Phis_{ij} z_j
                        -2Z_{ij}\Ps_j - Z_{jk} \Ps_{jki} -z_j Z_{ik} \Om_{jk},\ \\
  \Ps'_{ijk}  &\mapsto& \Ps'_{ijk} + 2z_{[k}\Phi_{j]i} + 2z_i\Phia_{jk} + z_l \Phi_{lijk}
                        + 2z_iz_{[k} \Ps_{j]} + 2z_lz_{[k} \Ps_{j]li} + Z_{il}\Ps_{ljk}\nn\\
              &       & + 2Z_{il} z_{[k}\Om_{j]l},\\
  \Om'_{ij}   &\mapsto& \Om'_{ij}-2z_{(j}\Ps'_{i)}+2z_k\Ps'_{(i|k|j)}
                        +2Z_{(i|k}\Phi_{k|j)}
                        +z_iz_j\Phi-4z_kz_{(i}\Phi^A_{j)k} + z_kz_l \Phi_{kilj}\nn\\
              &       & + 2z_{(i}Z_{j)k}\Ps_k + 2z_l Z_{(i|k}\Ps_{kl|j)} +Z_{ik}Z_{jl}\Om_{kl}.
\end{eqnarray}

\end{document}